\documentclass[aps,prb,showpacs,onecolumn,groupedaddress]{revtex4}
\usepackage{bm}
\usepackage{graphics}
\begin{document}

\title{Glassy phases in layered Ising magnets with random interlayer exchange}

\author{P. N. Timonin}
\email{timonin@aaanet.ru}
\affiliation{Physics Research Institute at Rostov State University
344090, Rostov - on - Don, Russia}

\date{\today}

\begin{abstract}
Thermodynamics of layered Ising magnet with the infinite-range ferromagnetic intralayer interaction and random exchange between nearest layers is considered. The case of zero average interlayer exchange is studied in detail. The inequilibrium thermodynamic potential is obtained near transition point and existence of numerous metastable states is shown. The thermodynamic properties of crystal in the simple periodic states are described. It is established that the evolution of the equilibrium magnetic state with growth of random exchange proceed over infinite series of first order transitions at which the number of magnetized layers and distance between them jump as well as homogeneous magnetization and susceptibility.
\end{abstract}


\maketitle 

The complete or partial disorder of magnetic structure has been found in many layered magnets and to describe it the term 'spin glass' has often been used \cite{1,2,3,4,5,6}. The other spin glass features were also observed in layered magnets such as the difference of 'field-cooled' and 'zero-field-cooled' parameters \cite{4,6}, smeared peak and low-frequency dispersion of magnetic susceptibility \cite{4,5,6}. The glassy properties do not exist only in the crystalline solid solutions, in which they obviousely result from magnetic exchange fluctuations caused by ion substitutions. Some crystals allow significant deviation from stechiometry, such as $SrCoO_{3-\delta}$ \cite{4,5}, in others, the disorder is their intrinsic property as in $LiNiO_2$ - here $Li$ and $Ni$ ions can easily exchange positions as they have almost equal sizes \cite{1}. Also the nominally pure crystals has always some impurities and defects.

	The simplest variant of disorder is the intercalation of impurities into the interlayer space, which does not modify the intralayer magnetic interactions. Similar model explains the glassy features of $LiNiO_2$ - small concentration of $Li$ ions in the ferromagnetic layers containing $Ni$ does not essentially affect them, while the presence of $Ni$ ions in $Li$ layers results in appearance of local ferromagnetic interlayer exchange on the background of small antiferromagnetic one \cite{1}. In this case one may hope to get comparatively simple description of disorder effects, especially when radius of magnetic intralayer interaction is much greater than a lattice parameter. Then intralayer ordering can be considered in the mean-field approximation everywhere except a narrow vicinity of transition point. Here the interlayer exchange fluctuations are averaged over intralayer interaction range, which can be put infinite in this mean-field region. Hence, we get a quasi one-dimensional situation with some average interlayer interaction \cite{7}. Then theoretical description becomes simpler than in case of $3d$ disorder and one may try to get some analytical results on the thermodynamics of such random layered magnet, preserving some specific features of spin glasses, particularly, the presence of multiple metastable states. The origin and properties of such states is a central issue of the theory of random magnets, so the study of such model could give not only approximate description of thermodynamics of real crystals but also shed some light on the mechanism of emergence of multiple metastable states responsible for the inergodic behavior of spin glasses.
	
Here we consider the simplest model of Ising magnet having infinite-range ferromagnetic intralayer interaction and random exchange between spins in neighboring layers. It allows to obtain rather easily an effective thermodynamic potential for the layers' magnetizations near transition point. Therefore, we can make sufficiently complete study of the origin and the properties of various metastable states for the case of zero average interlayer exchange. Further we establish the existence of a number of equilibrium spin-glass phases for different values of random exchange fluctuations and describe phase transitions between them.

\section{Model and formalism }
The Hamiltonian of the model is
\begin{equation}
{\cal H} =  - \frac{J}{{2M}}\sum\limits_{n = 1}^N {\left( {\sum\limits_{m = 1}^M {S_{m,n} } } \right)^2  - } \sum\limits_{m = 1}^M {\sum\limits_{n = 1}^N {\tilde J_{m,n} S_{m,n} S_{m,n + 1} } } -\sum\limits_{m = 1}^M {\sum\limits_{n = 1}^N {S_{m,n} H_n } } 
\label{eq:1}
\end{equation}
Here $S_{m,n}  =  \pm 1$ are Ising spins, index $m$ numerates the lattice sites in the layer from 1 to $M$, $n = 1,2,...,N $ is the layer number, $J > 0$ is ferromagnetic intralayer exchange, $\tilde J_{m,n} $ describes random interlayer interaction and $H_n $ is magnetic field, which is constant inside a layer. We assume all $\tilde J_{m,n} $ to have the same distribution function.

For the Hamiltonian in Eq. (\ref{eq:1}) one can get the averaged over $\tilde J_{m,n} $ value of the thermodynamic potential (per site in a layer)
\[
 M\beta \bar F = -\left\langle {\ln Tr\exp \left( { - \beta {\cal H}} \right)} \right\rangle _{\tilde J}.
\]
Here $\beta  = 1/T$ is inverse temperature, $Tr$ means the sum over all spin configurations and the angle parenthesis with index  $\tilde J$  below designate the average over random $\tilde J_{m,n} $. 

The Hubbard-Stratonovich transformation of the ferromagnetic term in Eq. (\ref{eq:1}) gives
\begin{equation}
M\beta \bar F = -\left\langle {\ln \left\{ {\int {d{\bf f}} \exp \left[ { - M\beta \tilde F\left( {\bf f} \right)} \right]} \right\}} \right\rangle _{\tilde J}, \label{eq:2}
\end{equation}
\begin{equation}
\beta \tilde F\left( {\bf f} \right) = \frac{1}{{2K}}\sum\limits_{n = 1}^N {f_n^2 }  - \frac{1}{M}\sum\limits_{m = 1}^M {\ln \left\{ {Tr\exp \sum\limits_{n = 1}^N {\left[ {\tilde K_{m,n} S_{m,n} S_{m,n + 1}  + \left( {h_n  + f_n } \right)S_{m,n} } \right]} } \right\}} ,
\label{eq:3}
\end{equation}
 where $h_n  = \beta H_n $  , $K = \beta J$ and $\tilde K_{m,n}  = \beta \tilde J_{m,n} $

As the logarithms in the sum in Eq. (\ref{eq:3}) are independent random quantities the averaging of Eq. (\ref{eq:2}) over random interlayer exchange  reduces to the averaging of this large sum when $M \to \infty $. Thus we get
\begin{eqnarray}
M\beta \bar F = -\ln \left\{ {\int {d{\bf f}} \exp \left[ { - M\beta F\left( {\bf f} \right)} \right]} \right\}, \label{eq:4}\\                               \beta F\left( {\bf f} \right) = \frac{1}{{2K}}\sum\limits_{n = 1}^N {f_n^2 }  - \left\langle {\ln \left\{ {Tr\exp \sum\limits_{n = 1}^N {\left[ {\tilde K_n S_n S_{n + 1}  + \left( {h_n  + f_n } \right)S_n } \right]} } \right\}} \right\rangle _{\tilde J}.\label{eq:5}
\end{eqnarray}

In Eq. (\ref{eq:5}) index $m$ is omitted at spins and random exchange integral because the contributions from all sites in the layers are identical. Obtaining the inequilibrium thermodynamic potential in Eq. (\ref{eq:5}), one can describe all thermodynamics of the model. Indeed, at $M \to \infty $ we have
\begin{equation}
\bar F \approx \mathop {\min }\limits_{\bf f} \left[ {F\left( {\bf f} \right)} \right] - TS_{conf},
\label{eq:6}
\end{equation}
where $S_{conf} $ the is configurational entropy defined via the logarithm of the number $W$ of points $\bf f $ at which the minimum of $ F\left( {\bf f} \right)$ is attained 
\[
S_{conf}  = M^{ - 1} \ln W.
\]
For the layer magnetization
\[
m_n  \equiv M^{ - 1} \sum\limits_{m = 1}^M {\left\langle {S_{m,n} } \right\rangle _{T,\tilde J} }  =  - \frac{{\partial \beta \bar F}}{{\partial h_n }},
\]
we get from Eqs. (\ref{eq:4}, \ref{eq:5})
\begin{equation}
m_n  =  - \left\langle {\frac{{\partial \beta F\left( {\bf f} \right)}}{{\partial h_n }}} \right\rangle _f  = \left\langle {K^{ - 1} f_n  - \frac{{\partial \beta F\left( {\bf f} \right)}}{{\partial f_n }}} \right\rangle _f  \approx K^{ - 1} \left\langle {f_n } \right\rangle _f ,
\label{eq:7}
\end{equation}
and for the correlator defining the inhomogeneous susceptipility
\[
C_{n,n'}  \equiv M^{ - 1} \sum\limits_{m,m' = 1}^M {\left\langle {\left\langle {S_{m,n} S_{m',n'} } \right\rangle _T  - \left\langle {S_{m,n} } \right\rangle _T \left\langle {S_{m',n'} } \right\rangle _T } \right\rangle _{\tilde J} }  =  - \frac{{\partial ^2 \beta \bar F}}{{\partial h_n \partial h_{n'} }},
\]
we have
\begin{equation}
C_{n,n'}  = MK^{ - 2} \left[ {\left\langle {f_n f_{n'} } \right\rangle _f  - \left\langle {f_n } \right\rangle _f \left\langle {f_{n'} } \right\rangle _f } \right] \approx K^{ - 2} \left. {\left[ {\frac{{\partial ^2 \beta F\left( {\bf f} \right)}}{{\partial {\bf f}\partial {\bf f}}}} \right]_{n,n',}^{ - 1} } \right|_{{\bf f} = \left\langle {\bf f} \right\rangle } .
\label{eq:8}
\end{equation}
The second term in Eq. (\ref{eq:5}) contains under the logarithm the partition function of Ising chain with random exchange in inhomogeneous field. It can be represented as expansion in powers of $\mu _n  \equiv \tanh\left( {h_n  + f_n } \right)$ and one can express the potential in Eq. (\ref{eq:5}) as  a function of these variables
\begin{eqnarray}
 \beta F\left( \mu  \right) =  - \left( {N - 1} \right)\left\langle {\ln \left( {2\cosh\tilde K} \right)} \right\rangle _{\tilde J}  + \frac{1}{2}\sum\limits_{n = 1}^N {\left[ {\left( {\tanh^{-1}(\mu _n)  - h_n } \right)^2 /K + \ln \left( {1 - \mu _n^2 } \right)} \right]}  - \nonumber \\ 
  - \left\langle {\ln \left( {1 + \sum\limits_{k = 1}^{\left[ {N/2} \right]} {\sum\limits_{n_1  < n_2  < ... < n_{2k} } {G_{n_1 ,n_2 ,...,n_{2k} } \prod\limits_{l = 1}^{2k} {\mu _{n_l } } } } } \right)} \right\rangle _{\tilde J},  \label{eq:9} 
\end{eqnarray}
\[
G_{n_1 ,n_2 ,...,n_{2k} }  = \prod\limits_{j = 1}^k {G_{n_{2j - 1} ,n_{2j} } }, 
\qquad
G_{n,n'}  = \prod\limits_{i = n}^{n'} {\tanh\left( {\tilde K_i } \right)} 
\]

To perform the averaging in the last term of Eq. (\ref{eq:9}) is generally a hard task, it is much simpler to study the vicinity of transition point and the case of small or zero external fields. Then we can leave in Eq. (\ref{eq:9}) only the lowest powers in its expansion in small $\mu_n$ and $h_n$. Further we consider just this case.
\section{Thermodynamics near transition point}
Up to the forth order in $\mu_n$ and for $\left| {h_n } \right| \le \left| {\mu _n^3 } \right|$  we get from Eq. (\ref{eq:9})
\begin{eqnarray}
\beta F\left( {\bf \mu } \right) =  - \left( {N - 1} \right)\left\langle {\ln \left( {2\cosh\tilde K} \right)} \right\rangle _{\tilde J}  + \frac{{{\bf h}^2 }}{{2K}} - \frac{{{\bf h\mu }}}{K} + \frac{1}{2}\sum\limits_{n,n'} {\mu _n \left( {K^{ - 1} \delta _{n,n'}  - g_{n - n'}^{\left( 1 \right)} } \right)\mu _{n'} }  + \left( {\frac{1}{{3K}} - \frac{1}{4}} \right)\sum\limits_n {\mu _n^4 } +\nonumber \\
\frac{1}{2}\sum\limits_{i < j} {\mu _i^2 g_{i - j}^{\left( 2 \right)} \mu _j^2 }  + \sum\limits_{i < j < k} {\left( {\mu _i^2 g_{i - j}^{\left( 2 \right)} \mu _j g_{j - k}^{\left( 1 \right)} \mu _k  + \mu _i g_{i - j}^{\left( 1 \right)} \mu _j g_{j - k}^{\left( 2 \right)} \mu _k^2 } \right)}  + 2\sum\limits_{i < j < k < l} {\mu _i g_{i - j}^{\left( 1 \right)} \mu _j g_{j - k}^{\left( 2 \right)} \mu _k g_{k - l}^{\left( 1 \right)} \mu _l } \label{eq:10}
\end{eqnarray}
where $g_n^{\left( 1 \right)}  = {\rm v}^{\left| {\rm n} \right|} $,
    $g_n^{\left( 2 \right)}  = u^{\left| {\rm n} \right|} $,
    ${\rm v} = \left\langle {\tanh\tilde K} \right\rangle _{\tilde J} $,
    $u = \left\langle {\tanh^2 \tilde K} \right\rangle _{\tilde J} $.
   
As   $\mu _n  \approx f_n /K$, the equilibrium (corresponding to the potential minimum) values of these variables give the layers' magnetizations, cf. Eq. (\ref{eq:7}). Apparently, in zero field the potential in Eq. (\ref{eq:10}) describes the second order phase transition at which spontaneous layers' magnetizations order ferromagnetically when $v>0$ or antiferromagnetically when $v<0$. The transition point in both cases is defined by the condition $K\sum\limits_{n =  - \infty }^\infty  {\left| {g_n^{\left( 1 \right)} } \right|}  = 1$. 

In general, the essentially nonlocal interaction of the forth order in Eq. (\ref{eq:10}) (its radius goes to infinity when $u \to 1$) allows to suppose that along with the minimum of the type $\mu _n  = \mu $ or $\mu _n  = \mu \left( { - 1} \right)^n $ the other minima could exist corresponding to some inhomogeneous ordering. Most probably, these minima would have larger values of the potential, thus defining the metastable states of a crystal.

Yet, in the presence of a field (especially inhomogeneous one) these states could become stable, i. e. there would be first order transitions from ferro- or antiferromagnetic state into various inhomogeneous ones depending on the strength and spatial variations the field applied. The dynamic phenomena in slow varying external fields would also have pecularities caused by the presence of inhomogeneous metastable states so their study in the ferro- or antiferromagnetic cases are very important for the description of the observed glassy properties of layered magnets. 
The interesting features are also revealed in the numeric study of temperature dependence of the magnetization in the case of random ferromagnetic excange \cite{7} - it shows visible oscillations growing while temperature is decreasing below the transition point. Nature of this phenomenon is not clear now.

However, the case of pure glass ordering, realized when the distribution function of interlayer exchange is symmetric, seems to be the most interesting one. Then $v = 0$ and the potential in Eq. (\ref{eq:10}) at $h_n=0$ becomes
\begin{equation}
\beta F\left( \mu  \right) =  - \left( {N - 1} \right)\left\langle {\ln \left( {2\cosh\tilde K} \right)} \right\rangle _{\tilde J}  + \frac{\tau }{2}\mu ^2  - \frac{1}{6}\sum\limits_n {\mu _n^4 }  + \frac{1}{4}\sum\limits_{n,n'} {\mu _n^2 u^{\left| {n - n'} \right|} \mu _{n'}^2 } 
\label{eq:11}
\end{equation}
where $\tau  \equiv K^{ - 1}  - 1 = \left( {T - J} \right)/J$,  $\tau \ll  1$.
In spite of apparent simplicity of Eq. (\ref{eq:11}) the determination of spontaneous magnetizations $\mu_n$ appearing at $\tau  < 0$ is rather complicated. The equation for the $F\left( \mu  \right)$ extrema
\begin{equation}
\mu _n \left( {\tau  - \frac{2}{3}\mu _n^2  + \sum\limits_l {u^{\left| {n - l} \right|} \mu _l^2 } } \right) = 0,
\label{eq:12}
\end{equation}
has in general case a number of solutions, among which one should choose the stable ones (i. e. corresponding to local minima) having the positively definite matrix of second order derivatives (Hessian)
\begin{equation}
\beta F''_{n,n'}  = \delta _{n,n'} \left( {\tau  - 2\mu _n^2  + \sum\limits_l {u^{\left| {n - l} \right|} \mu _l^2 } } \right) + 2\mu _n u^{\left| {n - n'} \right|} \mu _{n'}.                                                            \label{eq:13}
\end{equation}

The form of stable solutions of Eq. (\ref{eq:12}) depends essentially on the $u$ value. Indeed, in the trivial case $u = 0$ they are of the form $\mu _n  =  \pm \mu $, i. e. all layers have the same magnetizations' magnitudes while their orientations are arbitrary, so there are $2^N$ solutions. This degeneracy of the potential minima is the consequence of its symmetry with respect to the sign change of every $\mu _n $. In the other extreme when random interactions are infinite, $u \to 1$, the stable states have only one (arbitrary) layer with nonzero magnetization, thus there are $2N$ solutions (with different $\mu _n  \ne 0$ signs).
Our task is to find out which states are realized at the intermediate $u$ values and how the evolution of the magnetic state of crystal proceeds when disorder grows.

It follows from Eq. (\ref{eq:12}) that general form of its solution is
\begin{equation}
\mu _n  =  \pm \vartheta _n \sqrt {\left| \tau  \right|q_n }
\label{eq:14}
\end{equation}
The variables $\vartheta _n $ can take two values, $0$ and $1$, $\vartheta _n  = \left\{ {0,1} \right\}$. Choosing the definite vector $\bf \vartheta $, one can find $\bf q$ from Eq. (\ref{eq:12})
 \begin{eqnarray}
{\bf q} = \hat D^{ - 1} \vartheta , \label{eq:15}\\
D_{n,n'}  = \vartheta _n u^{\left| {n - n'} \right|} \vartheta _{n'}  - \frac{2}{3}\delta _{n,n'}.\label{eq:16}
\end{eqnarray}
Let us note that variable $\left| \tau  \right|{\bf q}$ has the meaning of the (inhomogeneous) Edwards-Anderson order parameter \cite{8}. According to Eq. (\ref{eq:14}) the choice of $\bf \vartheta $ for a given $u$ is limited by the condition $q_n  > 0$. Also the positive definiteness is necessary for Hessian, Eq. (\ref{eq:13}), or, equivalently, for the correlator, cf. Eq. (\ref{eq:8}), which has the following form on the solutions considered
\begin{eqnarray}
C_{n,n'}  = \left| \tau  \right|^{ - 1} \left[ {\vartheta _n \left( {\hat D^{ - 1} } \right)_{n,n'} \vartheta _{n'} /2\sqrt {q_n q_{n'} }  + \delta _{n,n'} \left( {1 - \vartheta _n } \right)E_n^{ - 1} } \right], \label{eq:17}\\	
E_n  = \sum\limits_l {u^{\left| {n - l} \right|} q_l \vartheta _l }  - 1 .
\label{eq:18}
\end{eqnarray}
For this the validity of the following conditions are sufficient 
\begin{eqnarray}
E_n  > 0
\qquad \text{at}
\qquad
\vartheta _n  = 0, \label{19}\\	
\vartheta _n D_{n,n'} \vartheta _{n'}  > 0. \label{eq:20}
\end{eqnarray}
The last inequality designates symbolically the positive defineteness of the matrix $\hat D$ on the subspace where $\vartheta _n  = 1$.

The corresponding values of the equlibrium thermodynamic potential are
\[
\beta \bar F =  - \left( {N - 1} \right)\left\langle {\ln \left( {2\cosh\tilde K} \right)} \right\rangle _{\tilde J}  + \frac{\tau }{4}\mu ^2  - S_{conf} = 
 - \left( {N - 1} \right)\left\langle {\ln \left( {2\cosh\tilde K} \right)} \right\rangle _{\tilde J}  - \frac{{\tau ^2 }}{4}\vartheta \hat D^{ - 1} \vartheta  - M^{ - 1} \sum\limits_n {\vartheta _n \ln 2}.
\]

\section{Periodic states}
To find among $2^N$ vectors $\bf \vartheta $ those obeying the above conditions at a given $u$ is rather difficult task. So we consider first only periodic states having some period $L$. Introducing the layers' numeration of the form $n = rL + s$ with integer $r$ and $s = 1,2,...,L$ we will study the solutions such that $\vartheta _{rL + s}  = \vartheta _s $. Apparently, the $L$-dimensional vectors $\vartheta _s $, which differ by a cyclic permutations of components, describe the same solution. Then it is convenient to consider $\vartheta _s $ as defined on the equidistance points of a circle to show visually that the rotation of $\vartheta _s $ on a circle do not change the state.

Let us also note that the signs of $\mu _n $ in the states considered should not be necessarily periodic as they do not enter the stability conditions and the equilibrium potential value. So further results refer to all $2^{\left[ {N/L} \right]P} $ states differing by layers' magnetization orientations. Here $P = \sum\limits_{s = 1}^L {\vartheta _s } $ is  the number of magnetized layers in a period and the quadrangle parentheses denote the integer part of a number.

For such periodic states we have
\[
D_{n,n'}  = D_{s,s'} \left( {r - r'} \right) = \vartheta _s \vartheta _{s'} u^{L\left| {r - r'} \right| + \left( {s - s'} \right)sign\left( {r - r'} \right)}  - \frac{2}{3}\delta _{r,r'} \delta _{s,s'} 
\]
As the matrix $\hat D$ depends only on $r - r'$ it can be diagonalized over this indexes using the unitary transformation $\Omega _{k,r}  = \left( {L/N} \right)^{1/2} \exp ikr$ in which we assume that $k$ belong to the first Brillouin zone,  $\left| k \right| < \pi $. The result is
\begin{equation}
D_{s,s'} \left( k \right) = \vartheta _s \vartheta _{s'} \left( {u^{\left| {s - s'} \right|}  + \frac{{u^{s - s'} }}{{u^{ - L} e^{ - ik}  - 1}} + \frac{{u^{s' - s} }}{{u^{ - L} e^{ik}  - 1}}} \right) - \frac{2}{3}\delta _{s,s'} 
\label{eq:21}
\end{equation}

Now the stability condition for the matrix $\hat D$, Eq. (\ref{eq:20}), means that $D_{s,s'} \left( k \right)$ must be positively defined on subspace where $\vartheta _s  = 1$ at all $\left| k \right| < \pi $.
Periodic solutions for $\bf q$ can be expressed via  $D_{s,s'} \left( k \right)$ at $k = 0$,
\begin{equation}
q_{rL + s}  = q_s  = \left[ {\hat D\left( 0 \right)^{ - 1} } \right]_{s,s'} \vartheta _{s'} 
\label{eq:22}
\end{equation}
the same is true for $E_s$
 \begin{equation}
E_{rL + s}  = E_s  = \hat D_{s,s'} \left( 0 \right)q_{s'} \vartheta _{s'}  - 1
\label{eq:23}
\end{equation}
Diagonalization of the correlator over $r - r'$ gives
\begin{equation}
C_{s,s} \left( k \right) = \left| \tau  \right|^{ - 1} \left\{ {\vartheta _s \left[ {\hat D\left( k \right)^{ - 1} } \right]_{s,s'} \vartheta _{s'} /2\sqrt {q_s q_{s'} }  + \delta _{s,s'} \left( {1 - \vartheta _s } \right)E_s^{ - 1} } \right\}, 
\label{eq:24}
\end{equation}
and for the equilibrium potential we have
\begin{equation}
\beta \bar F/N \approx  - \left\langle {\ln \left( {2\cosh\tilde K} \right)} \right\rangle _{\tilde J}  - L^{ - 1} \left[ {\frac{{\tau ^2 }}{4}\vartheta \hat D\left( 0 \right)^{ - 1} \vartheta  + {\rm M}^{ - 1} P\ln 2} \right] .
\label{eq:25}
\end{equation}

The last term in Eq. (\ref{eq:25}) proportional to $P$ is  the contribution of the configurational entropy appearing due to degeneracy of the potential minima as there are $2^{\left[ {N/L} \right]P}$ states with equal potentials differing by the signs of layers' magnetizations. It should be taken into account when $P$ is of order $M$. As $P \le N$, this could only happen if $N \approx M$, i. e. in thin whiskers with diameter of order $\sqrt M $ much smaller than the length $N$.  Yet we should note that this degeneracy are present only in the idealized situation when an external field is exactly zero while in real experiments there always is some small inhomogeneous field breaking this degeneracy. Further we assume the presence of such infinitesimal field and drop the contribution from the configurational entropy irrespective of the sample geometry.

\section{Thermodynamics and stability of periodic states}
Let us consider the simplest periodic states for which analytical results can be obtained. Thus, for the states with $P=1$ ($\vartheta _L  = 1,{\rm    }\vartheta _s  = 0,{\rm   }s = 1,...,L - 1$) the matrix $\hat D$ in Eq. (\ref{eq:21}) becomes just the function of $k$,
\begin{equation}
D\left( k \right) = \frac{1}{3}\frac{{1 + 4u^L \cos k - 5u^{2L} }}{{1 - 2u^L \cos k + u^{2L} }}
\label{eq:26}
\end{equation}
and we have
\begin{eqnarray}
q_L  \equiv q = D\left( 0 \right)^{ - 1}  = 3\frac{{1 - u^L }}{{1 + 5u^L }},\label{eq:27}\\
E_s  = 3\frac{{u^s  + u^{L - s} }}{{1 + 5u^L }} - 1, 
\qquad
s = 1,...,L - 1\label{eq:28}
\end{eqnarray}

Substitution of these expressions into Eq. (\ref{eq:24}) gives the formulae for the correlator which allows to find, in particular, the expression for the homogeneous susceptibility,

\begin{equation}
\beta \chi  = L^{ - 1} \sum\limits_{s,s'}^L {C_{s,s'} \left( 0 \right)}  = \left( {1 + 2\sum\limits_{s = 1}^{L - 1} {E_s^{ - 1} } } \right)/\left( {2\left| \tau  \right|L} \right)
\label{eq:29}
\end{equation}
It follows from Eqs. (\ref{eq:19}, \ref{eq:20}) that the stability region of these states is defined by the conditions $E_{\left[ {L/2} \right]}  > 0$, $D\left( \pi  \right) > 0$. The first of them defines the lower bound for $u$ values and the last one defines the upper bound. For the even $L$ we have
\begin{equation}
5^{ - 2}  < u^L  < 5^{ - 1} 
 \label{eq:30}
\end{equation}
For small odd $L$ the lower bound is slightly higher, but while $L$ grows it soon comes to $1/25$.

Introducing the variable
\begin{equation}
R\left( u \right) =  - \frac{{\ln 5}}{{\ln u}},	
\label{eq:31}
\end{equation}
having a meaning of effective radius of interlayer exchange, the stability condition in Eq. (\ref{eq:30}) can be represented in the form
\begin{equation}
1 < L/R < 2.
\label{eq:32}
\end{equation}
This representation demonstrates explicitly that period $L$ (the distance between magnetized layers) cannot significantly differ from the radius of interlayer exchange.
The equilibrium potential of the states considered is
\begin{equation}
\beta \bar F/N \approx  - \left\langle {\ln \left( {2\cosh\tilde K} \right)} \right\rangle _{\tilde J}  - \tau ^2 q/4L.	
\label{eq:33}
\end{equation}

Now we turn to the states with $P=2$. Let $\vartheta _L  = \vartheta _{L'}  = 1$, while the rest $\vartheta _s  = 0$ and assume for definiteness $L' \le L/2$. Then
\begin{eqnarray}
q_L  = q_{L'}  \equiv q' = \frac{{3\left( {1 - u^L } \right)}}{{1 + 3u^{L'}  + 3u^{L - L'}  + 5u^L }}
\label{eq:34}\\
\beta \bar F/N \approx  - \left\langle {\ln \left( {2\cosh\tilde K} \right)} \right\rangle _{\tilde J}  - \tau ^2 q'/2L
\label{eq:35}
\end{eqnarray}
For these states to be stable the conditions $E_{\left[ {\left( {L + L'} \right)/2} \right]}  > 0$ and $D_{LL} \left( 0 \right) - D_{LL'} \left( 0 \right) > 0$ must be fulfilled or (for $L+L'$ even)
\begin{eqnarray}
6\left( u^{\left( {L - L'} \right)/2}  + u^{\left( {L + L'} \right)/2} \right) - 3\left( u^{L'}  + u^{L - L'} \right) - 5u^L  - 1 > 0 , \nonumber\\
1 + 5u^L  - 3\left( u^{L'}  + u^{L - L'} \right) > 0 \label{eq:36}	
\end{eqnarray}
Summing these inequalities we get the condition $u^{\left( {3L' - L} \right)/2}  < 1$ so they both can be satisfied only when $L' > L/3$ or
\begin{equation}
1 < \nu  < 2
\label{eq:37}	
\end{equation}
where $\nu  \equiv {\left({L - L'}\right)}/{{L'}}$ is the ratio of the maximal distance between magnetized layers and the minimal one.

The stability conditions in Eq. (\ref{eq:36}) can be presented in the form similar to that of Eq. (\ref{eq:32})
\begin{equation}
A'\left( \nu  \right) < L'/R < B'\left( \nu  \right).
\label{eq:38}	
\end{equation}
Numerical analysis of Eq. (\ref{eq:36}) gives the following approximate expressions
\[
A'\left( \nu  \right) \approx 0.77 + 0.23/\nu, 
\qquad
B'\left( \nu  \right) \approx  - 0.23 + 2.23/\nu 
\]
For $\nu  = 2$ the stability region reduces to a point while at $\nu  = 1$ ($L' = L/2$) it coincides with Eq. (\ref{eq:32}) because the corresponding states are identical to those having $P=1$ and $L_1  = L/2$. It is easy to see that these states with $P=1$ have the stability region ($L/4 < R < L/2$) which includes the whole interval of Eq. (\ref{eq:36})  $L'/B'\left( \nu  \right) < R < L'/A'\left( \nu  \right)$. They also have the lower potential values than the $P=2$ ones. The last statement is the consequence of the maximal value of $q'$, Eq. (\ref{eq:34}), at  $ L' = L/2$. Thus the states with $P=2$ are always metastable.

Analytical study of states with $P>2$ is very cumbersome, yet their main properties could be established using numerical methods and qualitative considerations following from the exact results for $P=1,2$. Thus one can deduce from the preceding results that at all $P$ stable states should have the distances between magnetized layers of the order $R$. Introducing the minimal and the maximal values of such distances (and assuming the points representing layers to lie on a circle)
\[
L_{\min }  = \mathop {\min }\limits_{s,s'} \left( {\left| {s - s'} \right|,L - \left| {s - s'} \right|} \right),
\qquad
{\rm     }L_{\max }  = \mathop {\max }\limits_{s,s'} \left( {\left| {s - s'} \right|,L - \left| {s - s'} \right|} \right)
\qquad
  \text{for}
\qquad
\vartheta _s \vartheta _{s'}  = 1
\]
one can get the upper bound for the parameter $\nu  = L_{\max } /L_{\min }$ of the stable states.

Indeed, one can see that the stability condition for the matrix $\hat D(0)$ in Eq. (\ref{eq:21}) implies that its maximal nondiagonal element (on the subspace where $\vartheta _s  = 1$) must be smaller than the diagonal ones. Considering states with $P\gg 1$ (and, hence, $L >>R$) we get from this the approximate necessary condition
\begin{equation}
3u^{L_{\min } }  < 1,
\label{eq:39}	
\end{equation}
while from the condition $E_{L_{\max } /2}  > 0$, assuming $L_{\max }\approx R $ and $q_s \approx  q_{max} =3$ in Eq. (\ref{eq:23}) we have approximately	
\begin{equation}
6u^{L_{\max } /2}  > 1
\label{eq:40}	
\end{equation}
From Eqs. (\ref{eq:39}, \ref{eq:40}) it follows that only those states are stable in finite $R$ interval which have
\[
\nu < \nu _c  = 2\frac{{\ln 6}}{{\ln 3}} \approx {\rm 3}{\rm .262}.
\]

\begin{figure}
\includegraphics{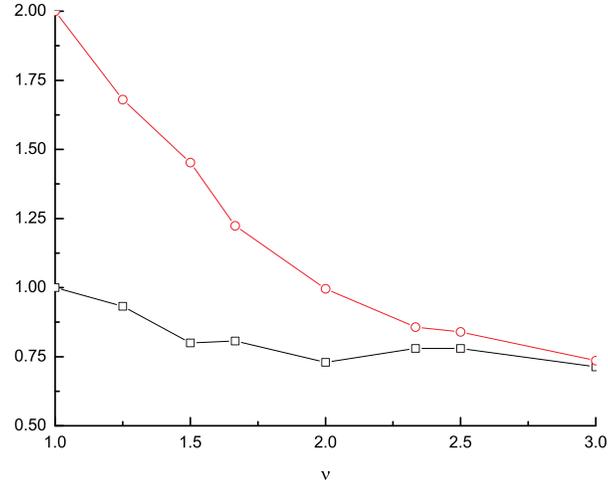}
\caption{\label{Fig.1}. The functions $A(\nu)$ (quadrangles) and $B(\nu)$ (circles) in Eq. (\ref{eq:41}) obtained in the numerical studies of stable states with $P>2$.}
\end{figure}

The largest $\nu$ value found in numerical studies of states with $5 \le L \le 30$ and $3 \le P \le 15$ is $3$. This corroborates the existence of upper bound $\nu _c  \approx 3$ for stable states. Numerical results show also that for $P>2$ there are the necessary stability conditions similar to those of Eq. (\ref{eq:38}),
\begin{equation}
A\left( \nu  \right) < L_{\min } /R < B\left( \nu  \right).
\label{eq:41}	
\end{equation}
The functions $A\left( \nu  \right)$  and $B\left( \nu  \right)$ obtained in the numerical studies are shown in Fig.\ref{Fig.1}. Note that the region between $A\left( \nu  \right)$ and $B\left( \nu  \right)$ includes the region defined by Eq. (\ref{eq:38}) and that Eq. (\ref{eq:41}) coincide with Eq. (\ref{eq:32}) at $\nu  \to 1$ as in this case all states become equivalent to those with $P=1$.
From Eq. (\ref{eq:41}) it follows that interval of $R$ values in which stable states with $P>1$ and $1 < \nu  < \nu _c  \approx 3$ exist,
\[
L_{\min } /B\left( \nu  \right) < R < L_{\min } /A\left( \nu  \right),
\]
belongs entirely to the stability region of states with $P=1$, Eq. (\ref{eq:32}), and some $L_1$ if $L_1$ obeys the condition
\[
L_{\min } /A\left( \nu  \right) < L_1  < 2L_{\min } /B\left( \nu  \right).
\]
Apparently such $L_1$ do exist and owing to $A\left( \nu  \right) \le 1$ and $\nu B\left( \nu  \right) \ge 2$ their values are confined between $L_{\min } $ and $L_{\max } $,
\[
L_{\min }  \le L_1  \le L_{\max }. 
\]
There are also states with $L_1$ in this interval and in its vicinity which cover partially the stability region of Eq. (\ref{eq:41}).

The mentioned above numerical studies show also that states with $P>1$ have always the higher thermodynamic potential than the coexisting with them $P=1$ states . The necessity of this can be shown analytically for the states with $\nu$ close to 1. Let us consider the states with $P>1$ and $L = PL_1 $ having equidistant magnetized layers, i. e. equivalent to the states with $P=1$ and $L_1  = L/P$. We can find the change of their potential, Eq. (\ref{eq:25}), under small variations of the positions of magnetized layers, $L_1 s \to L_1 s + \delta r_s $. The result is
\[
\beta \delta F/N = \left( {\tau q\ln u/2} \right)^2 \sum\limits_k {\left| {\delta r_k } \right|^2 \left[ {D\left( 0 \right) - D\left( k \right) - D\left( k \right)^{ - 1} \left| {D'\left( k \right)} \right|^2 } \right]} 
\]
Here $\delta r_k $ is Fourier-transform of  $\delta r_s $, $k = 2\pi l/P,{\rm     }l = 1,2,...,P$, $D\left( k \right)$ is given by Eq. (\ref{eq:26}) with $L=L_1$ and 
\[
D'\left( k \right) = \frac{{2iu_1 \sin k}}{{1 - 2u_1 \cos k + u_1^2 }},   
\qquad
u_1  \equiv u^{L_1 } .
\]
In the stability region $\left( {L_1 /2 < R < L_1 } \right)$, where $D\left( k \right) > 0$, the expression under the sum sign is positive for all $k$ so $\delta F > 0$ and states with $P>1$ and $\nu$ close to 1 have always the higher thermodynamic potential than the coexisting with them $P=1$ states.
 
We cannot give a rigorous proof of the metastability of states with $P>1$ at all $\nu  < \nu _c $. However, the numerical results allow to suggest that all states having different distances between magnetized layers have higher potential then that of the states with $P=1$ in which only one such distance ($L$) exists. The last means that nonperiodic states will also be metastable so equilibrium thermodynamic properties of the model are solely determined by the states with $P=1$. Further we consider the equilibrium thermodynamics and phase transition in the model in the framework of this supposition.

\section{Phase transitions at disorder strength changes}
The stability condition of $P=1$ states, Eq. (\ref{eq:32}), represented as $R < L < 2R $ shows that there are several stable states at all $ R  > 1 $. Hence one should find among them the states with the lowest potential, Eq. (\ref{eq:33}), to describe the equilibrium properties of a crystal. Considering the potential as continuous function of $L$ we find that it has minimum at
\[
L_c  = \lambda R,
\qquad
\lambda  \approx 1.471
\]
Thus the state with the lowest potential has $L$ closest to $L_c$. At
\[
R_L  = \lambda ^{ - 1} \left( {L + 1/2} \right) ,
\qquad
L > 1.
\]
there are two nearest and equidistant from $L_c$ periods ($L$ and $L+1$), so at that $R_L$ the jumps of the distance between magnetized layers occur manifesting the first order phase transitions.  At $R=1$ there is also phase transition between $L=1$ state (where all layers are magnetized) and $L=2$ state, this point being the stability boundary of both these phases.

\begin{figure}
\includegraphics{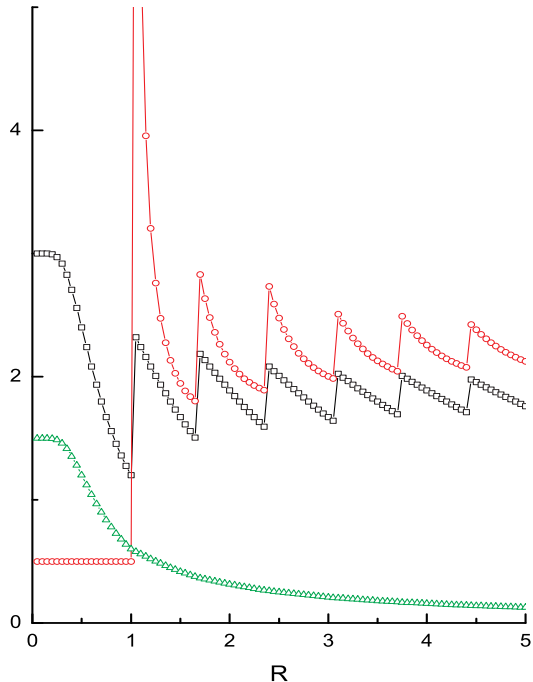}
\caption{\label{Fig.2}. The dependences of $q_{eq} $ (quadrangles), $\left| \tau  \right|\beta \chi _{eq} $ (circles) and $\delta C$ (triangles) on the effective interaction radius $R$. The jumps of first two parameters correspond to the phase transitions between the states with different distances between magnetized layers.}
\end{figure}

Thus the evolution of crystal state under the growth of disorder fluctuations (increasing of $u$ or $R$) proceed via infinite series of phase transitions accompanied by the jumps of the distance between magnetized layers as well as thermodynamic parameters of a crystal. To get this equilibrium parameters one should substitute in Eqs. (\ref{eq:27}, \ref{eq:29} , \ref{eq:33} ) the values of $L$ corresponding to a state with the lowest potential
\[
L_{eq} (R) = \vartheta \left( {1.5 - R} \right)\left[ {1 + \vartheta \left( {R - 1} \right)} \right] + \vartheta \left( {R - 1.5} \right)\sum\limits_{L = 2}^\infty  {L\vartheta \left[ {1 - 4\left( {\lambda R - L} \right)^2 } \right]} .
\]
This gives
\[
q_{eq} \left( R \right) = 3\frac{{1 - 5^{ - L_{eq} \left( R \right)/R} }}{{1 + 5^{1 - L_{eq} \left( R \right)/R} }},
\]
\[
\beta \chi _{eq} \left( R \right) = \left( {1 + 2\vartheta \left( {R - 1.5} \right)\sum\limits_{s = 1}^{L_{eq} \left( R \right) - 1} {E_{s,eq}^{ - 1} } } \right)/\left( {2\left| \tau  \right|L_{eq} \left( R \right)} \right).
\]
Let us remind that $q_{eq} \left( R \right)$ defines the spontaneous magnetization of layers $m_{eq} \left( R \right) = \sqrt {\left| \tau  \right|q_{eq} \left( R \right)} $. The behavior of  $q_{eq} \left( R \right)$ and the equilibrium homogeneous susceptibility is shown in Fig.\ref{Fig.2}.
Differentiating the equilibrium potential over the temperature  we get the magnetic contribution to the heat capacity appearing at $\tau  < 0$  ($T < J$) 
\[
\delta C = q_{eq} \left( R \right)/2L_{eq} \left( R \right).
\]
It has no jumps at the transitions being proportional to the equilibrium potential, see Fig.\ref{Fig.2}. 

It must be noted that these thermodynamic parameters are the same for all $2^{\left[ {N/L} \right]} $ states differing by the layers' magnetizations orientations while the total homogeneous magnetization is apparently different. It is determined by the infinitesimal field existing in the experiment and choosing the state in which $h_n \mu _n  > 0$ for all $n$. Then the overwhelming majority of these randomly chosen states would have zero or almost zero total magnetization. Indeed, the number of states with magnetization $m$ is
\[
N_m  = \left(\begin{array}{l}
 {\rm        }\left[ {N/L} \right] \\ 
 \left[ {N/L} \right]\left( {1 - m} \right)/2 \\ 
 \end{array} \right)  \approx 2^{\left[ {N/L} \right]} \exp  - \frac{{\left[ {N/L} \right]}}{2}\left[ {\left( {1 + m} \right)\ln \left( {1 + m} \right) + \left( {1 - m} \right)\ln \left( {1 - m} \right)} \right]
\]
It has narrow peak at $m = 0$ with the width of order $\left[ {N/L} \right]^{ - 1/2}$.

The results obtained so far can be slightly changed if potential expansion in $\mu$ will include higher order terms. Primarily, it concerns the vicinity of $R = 1$  point being the stability boundary of neighboring states. Here both phases becomes unstable with respect to the transition into $L=2$ phase of the type $\left( {\mu _1 ,\mu _2 } \right)$ which is absent in the forth-order approximation considered. Considering the higher order terms could result either in appearance of narrow (of order $\tau $) interval of existence of such phase with second order transition points at its boundaries, or in the overlapping of stability regions of the $L = 1$ and $L = 2$ phases \cite{9}. Also near the other transition points the states with $P >1$ could become stable so transitions could split into series of first order ones with long-periodic intermediate phases. Yet, all this cannot change the main conclusion about the  infinite series of transitions accompanying the growth of interlayer disorder.

Note also that in the immediate vicinity of transition point $T=J$ the present model has temperature dependencies of homogeneous magnetization and susceptibility characteristic to the ordinary mean-field ferromagnet. This is the consequence of the absence of disorder inside the layers undergoing purely ferromagnetic transitions. 

\section{Conclusions}
The present results give at least qualitative picture of magnetic structure and thermodynamic properties of layered Ising magnets having random interlayer exchange. Unlike the specific case considered here, really there always is some nonzero average interlayer interaction. It would split somewhat the potentials of degenerate glassy states, yet the growth of interlayer exchange fluctuations could still cause the changes of magnetic structure similar to ones described here. The study of such more realistic case can be done on the basis of present considerations. Qualitatively the presence of multiple almost degenerate metastable states with different total magnetizations would cause the inergodicity manifesting itself in a difference of 'field-cooled' and 'zero- field-cooled' parameters, specific hysteresis loop forms and long-time magnetic relaxation, which are really observed in many layered magnets \cite{1, 2, 3, 4, 5, 6}. 

It should be also noted that the existence of a number of glassy phases at different $u$ values means that transitions between them can also take place under temperature variations as $u$ depends on $T$. Probably the observed in [\onlinecite{7}] temperature oscillations of magnetization are the signatures of such transitions. One may also suppose that analogous yet more complex situation exists in case of short-range spin glasses, where  a continuous changes of magnetic structure under temperature variations has been supposed \cite{10}.

One may also consider the present model as the intermediate variant between the mean-field Sherrington - Kirkpatrick spin glass and the short-range Edwards-Anderson model \cite{8}. Combining the most of their essential features it allows for the first time to establish explicitly the existence of a number of metastable states and to study their properties which were not achieved in these classic models - there are only the estimates of a number of metastable states in the Sherrington - Kirkpatrick model \cite{11} and their phenomenological description in the short-range spin glass \cite{12}. Now it is hard to say to what extent the existence of metastable states relies on the presence of long-range interactions. One may only assume, taking into account the evidences of their appearance in real disordered magnets with dipole-dipole and magnetoelastic interactions, that sufficiently slow decaying interactions ensures the existence of numerous metastable states.
Anyway, the spin glass models without them have little chances to describe the inergodic phenomena observed in experiments.

\begin{acknowledgments}
This work was made under support from RFBR, grant 04-02-16228.
I gratefully acknowledge useful discussions with V. P. Sakhnenko, V. I. Torgashev, V. B. Shirokov, M. P. Ivliev and E. I. Gutlyanskii.
\end{acknowledgments}

\end{document}